\documentstyle[12pt]{article}
\begin{document}
\centerline{\Large\bf On the Dirac field in the Palatini form of 1/R gravity}
\vskip .7in
\centerline{Dan N. Vollick}
\centerline{Department of Physics and Astronomy}
\centerline{and}
\centerline{Department of Mathematics and Statistics}
\centerline{Okanagan University College}
\centerline{3333 University Way}
\centerline{Kelowna, B.C.}
\centerline{V1V 1V7}
\vskip .9in
\centerline{\bf\large Abstract}
\vskip 0.5in
In recent papers \cite{Vo1,Vo2} I have argued that the observed cosmological
acceleration can be accounted for by the inclusion of a $1/R$ term
in the gravitational action in the Palatini formalism. 
Subsequently, Flanagan \cite{Fl1,Fl2} argued that this theory is equivalent
to a scalar-tensor theory which produces corrections to the standard 
model that are ruled out experimentally. 
  
In this article I examine the Dirac field coupled to 1/R gravity. The Dirac
action contains the connection which was taken to be the Christoffel symbol,
not an independent quantity,
in the papers by Flanagan. Since the metric and connection are taken to be
independent in the Palatini approach it is natural to allow the connection
that appears in the Dirac action to be an independent quantity. This is the
approach that is taken in this paper.
The resulting theory is very different
and much more complicated than the one discussed in Flanagan's papers.
\vskip 0.5in
\newpage
Recently there have been several attempts to explain the observed
cosmological acceleration \cite{To1,Pe1,Ri1,Pe2,Be1,Ne1,Ha1,Ja1,La1,Me1} by including a $1/R$ term in the gravitational
Lagrangian. Capozziello et al. \cite{Cap1} and Carroll et al. 
\cite{Ca1} used a purely metric variation 
and I \cite{Vo1} used a Palatini approach to obtain the field equations.
In recent articles Flanagan \cite{Fl1,Fl2} 
has argued that the Palatini form of $1/R$ gravity
is in conflict with results from particle physics experiments.
  
In this article I examine the Dirac field coupled to 1/R gravity in the Palatini
formalism. In this approach the metric and the connection are taken to be
independent quantities. Thus, it makes sense to take the connection that
appears in the Dirac action to be independent. In his papers Flanagan 
has taken the metric and connection to be independent in the gravitational
part of the action but has taken the connection to be the usual Christoffel
symbol in the Dirac part of the action. This is, of course, 
mathematically consistent
but it does not seem natural within the Palatini formalism. In this article
I take the metric and connection to be independent in both parts of the action
and show that the resulting theory is very different
and much more complicated than the one discussed in Flanagan's papers.
The calculations in this article 
follow van Nieuwenhuizen's \cite{va1} derivation of the field equations 
for a gravitino coupled to gravity.
  
Consider a four dimensional manifold $M$ with metric $g_{\mu\nu}$ and tetrad
$e_{\;\;\mu}^{a}$.
The Riemann tensor is given by
\begin{equation}
R^{\;\;\;\;ab}_{\mu\nu}=\partial_{\mu}\omega_{\nu}^{\;\;ab}-\partial_{\nu}\omega_{\mu}
^{\;\;ab}+\omega_{\mu}^{\;\;ac}\omega_{\nu c}^{\;\;\;\;b}-\omega_{\nu}^{\;\;ac}
\omega_{\mu c}^{\;\;\;\;b}\; ,
\end{equation}
where $\omega_{\mu}^{\;\; ab}$ is the spin connection.
The first tetrad postulate is given by
\begin{equation}
\partial_{\mu}e^a_{\;\;\nu}+\omega_{\mu\;\;\; b}^{\;\;a}(e)e^b_{\;\;\nu}
-\Gamma_{\mu\nu}^{\alpha}(g)e^a_{\;\;\alpha}=0\; ,
\label{tetrad1}
\end{equation}
where
\begin{equation}
\omega_{\mu}^{\;\;ab}(e)=\frac{1}{2}e^{a\nu}\left[\partial_{\mu}e^b_{\;\;\nu}-\partial
_{\nu}e^b_{\;\;\mu}\right]-\frac{1}{2}e^{\nu b}\left[\partial_{\mu}e^a_{\;\;\nu}-\partial
_{\nu}e^a_{\;\;\mu}\right]-\frac{1}{2}e^{a\rho}e^{b\sigma}e^c_{\;\;\mu}\left[\partial
_{\rho}e_{c\sigma}-\partial_{\sigma}e_{c\rho}\right]\; ,
\label{connection}
\end{equation}
and $\Gamma^{\alpha}_{\mu\nu}(g)$ is the symmetric Christoeffel connection.
In a Palatini variation $\omega_{\mu}^
{\;\;ab}$ and $\Gamma^{\alpha}_{\mu\nu}$ are not considered as functions
of $e^a_{\;\;\mu}$ and $g_{\mu\nu}$. Thus, we introduce the second 
tetrad postulate
\begin{equation}
\partial_{\mu}e^a_{\;\;\nu}+\omega_{\mu\;\;\; b}^{\;\;a}e^b_{\;\;\nu}
-\Gamma_{\mu\nu}^{\alpha}e^a_{\;\;\alpha}=0.
\label{tetrad2}
\end{equation}
The gravitational Lagrangian is given by
\begin{equation}
L_G=\frac{e}{2\kappa^2}f(R)\; ,
\label{action}
\end{equation}
where 
\begin{equation}
R=e^{\;\;\mu}_ae^{\;\;\nu}_bR^{\;\;\;\;ab}_{\mu\nu}
\label{Ricci}
\end{equation}
and $e=det(e^a_{\;\;\mu})$. The matter Lagrangian will be taken to be
\begin{equation}
L_D=e\bar{\Psi}_e\left[i\gamma^{\mu}D_{\mu}-m\right]\Psi_e\; ,
\end{equation}
where $\gamma^{\mu}=e^{\;\;\mu}_a\gamma^a$, $\;\{\gamma^a\gamma^b\}=2\eta^{ab}$,
\begin{equation}
D_{\mu}\Psi_e=\partial_{\mu}\Psi_e+\frac{1}{2}\omega_{\mu}^{\;\;ab}\Sigma_{ab}\Psi_e
\end{equation}
and $\Sigma_{ab}=\frac{1}{4}[\gamma^a,\gamma^b]$ are the generators of 
the Dirac representation. Note that the Dirac Lagrangian depends on the 
spin connection.
Varying the action with respect to $\bar{\Psi}_e$ gives the Dirac equation
\begin{equation}
i\gamma^{\mu}D_{\mu}\Psi_e-m\Psi_e=0
\end{equation}
and varying the action with respect to the tetrad gives
\begin{equation}
f'(R)R_{\mu\nu}-\frac{1}{2}f(R)g_{\mu\nu}=-i\kappa^2
\bar{\Psi}_e\gamma_{\nu}D_{\mu}\Psi_e\; ,
\label{Einstein}
\end{equation} 
where
\begin{equation}
R_{\mu\nu}=R^{\;\;\;\;ab}_{\mu\beta}e_{a\nu}e^{\;\;\beta}_b\; .
\end{equation}
Note that the Ricci tensor and the
energy-momentum tensor of the Dirac field are not 
symmetric. A nonsymmetric energy-momentum tensor is not unusual 
in a Palatini-tetrad variational approach. For example, the 
energy-momentum tensor of the gravitino is not symmetric 
in supergravity theories (see \cite{va1}). 
    
Now consider varying the action with respect to the spin connection.
It will be convenient to use
\begin{equation}
R=\frac{1}{4e}\epsilon^{\mu\nu\alpha\beta}\epsilon_{abcd}e^a_{\;\;\mu}
e^b_{\;\;\nu}R_{\alpha\beta}^{\;\;\;\;cd}
\end{equation}
instead of (\ref{Ricci}). The variation of the gravitational 
Lagrangian gives
\begin{equation}
\delta L_G=\frac{1}{2\kappa^2}\epsilon^{\mu\nu\alpha\beta}\epsilon_{abcd}\left(
D_{\beta}\tilde{e}^a_{\;\;\mu}\right)\tilde{e}^b_{\;\;\nu}\delta\omega_{\alpha}^{\;\;cd}
\; ,
\label{grav}
\end{equation}
where
\begin{equation}
D_{\beta}\tilde{e}^a_{\;\;\mu}=\partial_{\beta}\tilde{e}^a_{\;\;\mu}+\omega_{\beta}^
{\;\;ab}\tilde{e}_{b\mu}\;\;\;\;\;\;
\textrm{and}\;\;\;\;\;\;
\tilde{e}^a_{\;\;\mu}=(f^{'})^{1/2}e^a_{\;\;\mu}\; .
\end{equation}
The variation of the Dirac Lagrangian gives
\begin{equation}
\delta L_D=\frac{1}{2}ie\bar{\Psi}_e\gamma^{\alpha}\Sigma_{cd}\Psi_e\delta\omega
_{\alpha}^{\;\;cd}\; .
\end{equation}
Now use the identities 
\begin{equation}
\Sigma_{cd}=-\frac{1}{2}\gamma_5\epsilon_{cdab}\Sigma^{ab}\; ,\;\;\;\;\;\;\;\;
\gamma^a=\frac{1}{6}\epsilon^{abcd}\gamma_5\gamma_b\gamma_c\gamma_d\; ,
\end{equation}
$(\gamma_5)^2=I$ and $\{\gamma_5,\gamma_{\mu}\}=0$ to write
the variation of the Dirac Lagrangian as 
\begin{equation}
\delta L_D=\frac{i}{24}\epsilon^{\mu\nu\alpha\beta}\epsilon_{abcd}\bar{\Psi}_e
\gamma_{\mu}\gamma_{\nu}\gamma_{\beta}\;\Sigma^{ab}\Psi_e\delta
\omega_{\alpha}^{\;\; cd}\; .
\label{Dirac}
\end{equation}
Comparison of (\ref{grav}) and (\ref{Dirac}) gives
\begin{equation}
\left(D_{[\beta}\tilde{e}^{[a}_{\;\;\mu}\right)\tilde{e}_{\;\;\nu]}^{b]}=\frac{i\kappa^2}{12}
\bar{\Psi}_e\gamma_{[\beta}\gamma_{\mu}\gamma_{\nu]}\Sigma^{ab}\Psi_e
\; .
\end{equation}
Now use 
\begin{equation}
\gamma_{[a}\gamma_{b}\gamma_{c]}=\epsilon_{abcd}\gamma_5\gamma^d
\end{equation}
to obtain
\begin{equation}
\left(D_{[\beta}\tilde{e}^{[a}_{\;\;\mu}\right)\tilde{e}_{\;\;\nu]}^{b]}=
\frac{i\kappa^2}{12}e\epsilon_{\beta\mu\nu\sigma}
\bar{\Psi}_e\gamma_5\gamma^{\sigma}\Sigma^{ab}\Psi_e\; .
\end{equation}
By contracting over $\tilde{e}_b^{\;\;\nu}$ and then over $\tilde{e}
_a^{\;\;\mu}$ and by using 
\begin{equation}
\gamma_{\mu}\Sigma_{ab}=\frac{1}{2}\left(e_{a\mu}\gamma_b-e_{b\mu}\gamma_a+e^c_{\;\;\mu}
\epsilon_{abcd}\gamma_5\gamma^d\right)
\end{equation}
it can be shown that
\begin{equation}
D_{\beta}\tilde{e}_{\;\;\mu}^a-D_{\mu}\tilde{e}^a_{\;\;\beta}=\frac{i\kappa^2}{4\sqrt{f^{'}}}
\bar{\Psi}_e\left[2e\gamma_5\epsilon_{\beta\mu\rho}^{\;\;\;\;\;\;a}\gamma^{\rho}-
\left(e^a_{\;\;\beta}\gamma_{\mu}-e^a_{\;\;\mu}\gamma_{\beta}\right)\right]\Psi_e\;.
\label{diff}
\end{equation}
From this and the second tetrad postulate it is easy to show that the spacetime
torsion is non vanishing.
Now define
\begin{equation}
\tilde{\omega}_{\mu}^{\;\;ab}=\omega_{\mu}^{\;\;ab}(\tilde{e})+K_{\mu}^{\;\;ab}
\label{cont}
\end{equation}
where $K_{\mu}^{\;\;ab}$ is the contorsion tensor, $\omega_{\mu}^
{\;\;ab}(\tilde{e})$ is given by (\ref{connection}) with 
$e^a_{\;\;\mu}$ replaced by $\tilde{e}^a_{\;\;\mu}$ and $\tilde{\omega}
_{\mu}^{\;\; ab}$ is the spin connection that appears in the second tetrad
postulate (\ref{tetrad2}) along with $\tilde{e}^a_{\;\;\mu}$.
From (\ref{diff})
and the first tetrad postulate we find
\begin{equation}
K_{\mu a\nu}-K_{\nu a\mu}=\frac{i\kappa^2}{4f^{'}}\bar{\Psi}_e\left[2\gamma_5
\epsilon_{\mu\nu\rho a}\gamma^{\rho}-\left(e_{a\mu}\gamma_{\nu}-
e_{a\nu}\gamma_{\mu}\right)\right]\Psi_e\; .
\end{equation}
Using the identity
\begin{equation}
K_{\mu a\nu}=\frac{1}{2}\left[(K_{\mu a\nu}-K_{\nu a\mu})+
(K_{a\mu\nu}-K_{\nu\mu a})+(K_{a\nu\mu}
-K_{\mu\nu a})\right]
\end{equation}
gives
\begin{equation}
K_{\mu}^{\;\;ab}=\frac{i\kappa^2}{4f^{'}}\bar{\Psi}_e\left[e
\gamma_5\epsilon_{\mu\;\;\;\;
\rho}^{\;\;ab}\gamma^{\rho}-\left(e^a_{\;\;\mu}\gamma^b-e^b_{\;\;\mu}
\gamma^a\right)\right]\Psi_e\; .
\label{K}
\end{equation}
Now from the first and second tetrad postulates we have
\begin{equation}
\tilde{\Gamma}^{\alpha}_{\mu\nu}=\Gamma^{\alpha}_{\mu\nu}(h)+K^{\;\;\alpha}_
{\mu\;\;\;\;\nu}\; ,
\end{equation}
where $h_{\mu\nu}=f^{'}g_{\mu\nu}$ is the metric associated with the tetrad
$\tilde{e}^a_{\;\;\mu}$ and $\tilde{\Gamma}^{\alpha}_{\mu\nu}$ is the connection
that appears in the second tetrad postulate (\ref{tetrad2}) along with 
$\tilde{e}_{\;\;\mu}^a$.
In terms of the original metric we have
\begin{equation}
\Gamma^{\alpha}_{\mu\nu}=\Gamma^{\alpha}_{\mu\nu}(g)+H^{\alpha}_{\mu\nu}\;
,
\end{equation}
where
\begin{equation}
H^{\alpha}_{\mu\nu}=\delta_{(\mu}^{\alpha}\nabla_{\nu)}\ln f^{'}-\frac{1}{2}
g_{\mu\nu}\nabla^{\alpha}\ln f^{'}+K^{\;\;\alpha}_{\mu\;\;\;\;\nu}
\label{H1}
\end{equation}
and $\nabla_{\mu}$ is defined with respect to the metric $g_{\mu\nu}$.

In his paper (see equation (19)) Flanagan argues that the connection is
always on-shell compatible with some metric $e^{2\chi}g_{\mu\nu}$ with
\begin{equation}
H^{\alpha}_{\mu\nu}=\delta_{(\mu}^{\alpha}\nabla_{\nu)}\chi-\frac{1}{2}
g_{\mu\nu}\nabla^{\alpha}\chi\; .
\label{H2}
\end{equation}
We see that this is no longer true if Fermions are present in the theory
(note that the $ln\chi$ that appears in his paper has been replaced by 
$2\chi$, as is appropriate for the metric $e^{2\chi}g_{\mu\nu}$).
In the tetrad formulation we find analogous results. Set
\begin{equation}
\omega_{\mu}^{\;\; ab}=\omega_{\mu}^{\;\; ab}(e)+h_{\mu}^{\;\; ab}
\end{equation}
and we find that
\begin{equation}
h_{\mu}^{\;\; ab}=\frac{1}{2}\left[e^a_{\;\;\mu}D^b\ln f^{'}-
e^b_{\;\;\mu} D^a\ln f^{'}\right]+K_{\mu}^{\;\; ab}\; .
\label{h}
\end{equation}
Instead of the ansatz (\ref{H2}) we might try
\begin{equation}
h_{\mu}^{\;\; ab}=\left[e^a_{\;\;\mu}D^b\chi-e^b_{\;\;\mu} D^a
\chi\right]+\frac{i\kappa^2}{4}e^{-2\chi}\bar{\Psi}_e\left[e\gamma_5
\epsilon_{\mu\;\;\;\;
\rho}^{\;\;ab}\gamma^{\rho}-\left(e^a_{\;\;\mu}\gamma^b-e^b_{\;\;\mu}
\gamma^a\right)\right]\Psi_e\; , 
\end{equation}
where all factors of $f^{'}(R)$ in
(\ref{h}) have been replaced by $e^{2\chi}$. It is important to note 
that the matter Lagrangian also
contains $\chi$ and it derivatives. This leads to a very different and much more complicated theory than the one discussed by Flanagan.
The next step in his argument involves transforming the action to the Einstein frame and showing that the 
resulting non minimally coupled terms are ruled out experimentally. 
Since I do not agree with this approach (see \cite{Vo2})
I will stop here and not proceed any further.
  
One final comment. Flanagan uses the well known equivalence theorems
to argue that the S matrix is invariant under nonlinear local field redefinitions
of $g_{\mu\nu}$ and $\Psi_e$. However, these theorems apply to fields on
a fixed Minkowski background spacetime and there is no guarantee that they
apply to the metric, which determines the spacetime structure. 
\section*{Acknowledgements}
I would like to thank \'Eanna Flanagan for helpful discussions concerning
the Palatini form of 1/R gravity.

\end{document}